\documentclass[aps,pra,amsmath,amssymb,twocolumn,showpacs,nofootinbib]{revtex4-1}

\usepackage{graphicx}
\usepackage{dcolumn}
\usepackage{bm}
\usepackage{amsmath}
\usepackage[normalem]{ulem}
\usepackage[flushleft]{threeparttable}
\usepackage{braket}
\usepackage{float}
\usepackage{microtype} 

\usepackage{amsmath,amsfonts,amsthm} 

\usepackage[svgnames]{xcolor} 

\usepackage[hang, small, labelfont=bf, up, textfont=it]{caption} 

\usepackage{booktabs} 

\usepackage{lastpage} 

\usepackage{graphicx} 

\usepackage[lofdepth,lotdepth]{subfig}

\usepackage{enumitem} 
\setlist{noitemsep} 

\usepackage[english]{babel}

\newcommand{\rom}[1]{\uppercase\expandafter{\romannumeral #1\relax}}

\begin{document}


\title{Continuous Monitoring of Energy in Quantum Open Systems}

\author{G. P. Martins}
  \email{gabrielll_63@hotmail.com}

\affiliation{ 
Departamento de F{\'i}sica, Universidade Federal de Minas Gerais, Belo Horizonte, MG, Brazil 
}%

\author{N. K. Bernardes}%

\affiliation{ 
Departamento de F{\'i}sica, Universidade Federal de Pernambuco, 50670-901 Recife, PE, Brazil
}%

\author{M. F. Santos}
\affiliation{%
Instituto de F{\'i}sica, Universidade Federal do Rio de Janeiro, Caixa Postal 68528, Rio de Janeiro, RJ 21941-972, Brazil
}%

\date{\today}

\begin{abstract}
We propose a method to continually monitor the energy of a quantum system. We show that by having some previous knowledge of the system's dynamics, but not all of it, one can use the measured energy to determine many other quantities, such as the work performed on the system, the heat exchanged between the system and a thermal reservoir, the time dependence of the Hamiltonian of the system as well as the total entropy produced by its dynamics. We have also analyzed how this method is dependent on the quality factor of the measurements employed. 
\end{abstract}

\maketitle

\section{Introduction}

The field of thermodynamics is one of the most interesting and peculiar of all physics. It has given us the knowledge to build thermal engines and thus to reach the level of development that is seen in today's world. Its study has allowed us to understand the notion of irreversible processes and the arrow of time\cite{lebowitz1993boltzmann}. While thermodynamical properties of macroscopic systems are widely understood, the focus on microscopic systems is relatively new. 



On the realm of stochastic thermodynamics, one iconic paper by Allicki has defined the form of the Laws of Thermodynamics for microscopic systems \cite{alicki1979quantum}. In this regime, the concept of thermodynamics is extended to a single stochastic trajectory of a microscopic system in phase space. In each realization of the same thermodynamical process, the system may undergo different trajectories and quantities such as work, heat and entropy production are no longer perfectly defined: they become trajectory dependent, and, therefore, also stochastic \cite{seifert2005entropy,sekimoto2010stochastic,jarzynski2011equalities}. In such cases, the Second Law of Thermodynamics is associated to the central Fluctuation Theorem (CT) \cite{evans2002fluctuation}, $\left\langle\exp\left(-\Delta S_{tot}\right)\right\rangle = 1$, where $\langle \rangle$ is to be understood as the average over many trajectories. Note that apparent violations of the second law may happen in any given trajectory \cite{evans1993probability,PhysRevE.50.1645,wang2002experimental}, although, on average, it still holds, as expected. Also note that, just as the thermodynamical law itself, the Central Fluctuation Theorem also has many formulations. An interesting example is the Jarzinski Equality \cite{jarzynski1997nonequilibrium} (JE) $\left\langle\exp\left(-W/k_BT\right)\right\rangle = \exp\left(-\Delta F/k_BT\right)$, which has been experimentally verified various times in recent years \cite{collin2005verification,toyabe2010experimental,douarche2005experimental,an2015experimental}.

Turning to the First Law of Thermodynamics, the definition of work, and how to measure it are no trivial tasks for microscopic quantum systems. The main reasons are the intrinsic exchange of energy between system and measurement apparatus and the also intrinsic random nature of the results of quantum measurements. Over the past decade, many such definitions were made \cite{skrzypczyk2014work,gallego2016thermodynamic,aaberg2013truly,roncaglia2014work,de2015measuring,allahverdyan2005fluctuations,jarzynskia2008nonequilibrium}, each better suited for a different situation. 
That said, whenever the dynamics and measurements do not involve coherences in the energy eigenbasis, thermodynamical quantities such as the exchange of work and heat as well as the entropy production are well defined and computable. That is the case here analysed. 

The goal of this paper is to propose a method to continually monitor the energy of a quantum system and to relate the monitored energy to its thermodynamics. In particular, we show that, by having some previous knowledge of the system's dynamics, one can use the measured energy to determine quantities such as the work performed on the system, the heat exchanged between the system and a thermal reservoir, the time dependence of the system's full Hamiltonian and the total entropy produced by its dynamics.

The paper is organized as follows: First, in section \rom{2} we review the protocol for measuring work in quantum systems proposed in \cite{roncaglia2014work, de2015measuring}. Second, in section \rom{3}, we transform it into a protocol to continually monitor the energy of a quantum system. In section \rom{4}, we analyse the effects of imperfect measures and define a quality factor for them. In Section \rom{5}, we study this quality factor in detail. In Section \rom{6} we apply our results to a two-level system subjected to a time dependent $\sigma_z$-type Hamiltonian. 
In Section \rom{7} we display and analyse simulations of quantum trajectories for the system. In section \rom{8}, we show how to combine the results of section \rom{6} in order to compute the total entropy produced by the dynamics, connecting the previous results to the Second Law of Thermodynamics. We also show that the measured value for the entropy production obeys an equation similar to the Jarzynski Equality with a correction term that tends to zero as the quality factor increases. Finally, section \rom{9} concludes the manuscript.



\section{The Roncaglia-Cerisola-Paz Protocol}

One task in thermodynamics is to evaluate the variation of internal energy of a system due to the exchange of work and heat with external agents, from which derives the first law of thermodynamics. In particular, let us consider the following problem: one wants to measure the variation of internal energy $\Delta U$ of a system that is driven from an initial Hamiltonian $H$, with eigenvalues equation $H \ket{\phi_n} = E_n \ket{\phi_n}$ at $t_0$, to a final Hamiltonian $\tilde{H}$, with eigenvalues equation $\tilde{H} \ket{\tilde{\phi}_n} = \tilde{E}_n \ket{\tilde{\phi}_n}$ at $t_f$, via an unitary evolution $V_E$. If the initial state of the system is $\rho_0$, the probability distribution $P(\Delta U)$ of obtaining $\Delta U$, through the measurement of its energy before and after the evolution, is given by
\begin{equation}
P(\Delta U) = \sum_{m,n}p_n p_{m,n}\delta(\Delta U-E_{m,n}), \label{pdist}
\end{equation}
where $p_n = \bra{\phi_n} \rho_0 \ket{\phi_n}$ is the probability to measure the system in $\ket{\phi_n}$  before the evolution, $p_{m,n} = \left| \bra{\tilde{\phi}_m} V_E \ket{\phi_n}\right| ^2$ is the probability to measure $\ket{\tilde{\phi}_m}$ at the end of the evolution, given that $\ket{\phi_n}$ was measured at the beginning, and $E_{m,n} = \tilde{E}_m -E_n$.

The protocol proposed in Ref.~\cite{roncaglia2014work} shows how to obtain the same probability distribution by making a single measure on an enlarged system. Before the system of interest $\mathcal{S}$ evolves, it interacts with an ancilla $\mathcal{A}$ that is initially in the vacuum state $\ket{0}$. The evolution of the composed system follows an interaction Hamiltonian $H_I$ given by
\begin{equation}
H_I = \dfrac{\lambda}{\hbar} H\otimes P_\mathcal{A},
\label{displ}
\end{equation}
where $H$ is the original Hamiltonian of the system $\mathcal{S}$, $P_{\mathcal{A}}$ is the (adimensional) generator of displacements of the ancilla $\mathcal{A}$, and $\lambda$ is a constant that parametrizes the $\mathcal{S} - {\mathcal{A}}$ interaction. If the system is originally in the state $\ket{\psi_0}$ and the total time of the interaction is equal to $\delta t_I$, this interaction produces the generally entangled system-ancilla state:
\begin{equation}
V_I \ket{\psi_0} \otimes \ket{0} = \sum_n \braket{\phi_n | \psi_0} \ket{\phi_n} \otimes \ket{\lambda t_I E_n}.
\end{equation}

Now, if the composed system $\mathcal{S+A}$ evolves via the following sequence of unitaries $V_{IEI} = \tilde{V}_I V_E V^{\dagger}_I$, its final state $\ket{\Psi_F} = V_{IEI}\ket{\psi_0} \otimes \ket{0}$ is given by
\begin{equation}
\ket{\Psi_F} = \sum_{m,n}\bra{\tilde{\phi}_m}V_E\ket{\phi_n} \braket{\phi_n | \psi_0} \ket{\tilde{\phi}_m}\otimes\ket{\lambda t_I E_{m,n}}. \label{psif}
\end{equation}
In this case, a measurement on the position of the ancilla results in the value $\Delta U$ with the same probability distribution of (\ref{pdist}), as long as the states of the ancilla are orthogonal.

If the evolution of the system is purely Hamiltonian, $\Delta U$ can be associated to the work $W$ performed on it. The protocol, however, is general and, when used on a system in contact with a thermal bath, it still provides the overall variation of internal energy. Since, in this case, the system can exchange heat with the bath, it can no longer be said that this variation was due solely to work performed on the system and the separation between heat and work may not be easily addressable. Nonetheless, as we show throughout the rest of the paper, there are particular cases where the continuous monitoring of $\Delta U$ can be used to differentiate between them and the full thermodynamics of the time evolution of the system is obtainable.


\section{Continuous Monitoring}
For simplicity, let us first consider the case where the system of interest is isolated, evolving by the usual Schr{\"o}dinger equation:
\begin{equation}
\dot{\rho} = -\dfrac{i}{\hbar}[\rho ,H].
\end{equation}
In this scenario, the most important condition in order 
to continuously monitor the energy exchanged by the system is that the total time of the measure $dt$, composed by the two interaction times $t_I$ plus the evolution time $t_E$ must be small enough so that $H(t)$ may be considered linear in $dt$. Also, the interaction time must be much smaller than the evolution time, so that $H(t)$ remains practically unchanged during the two interactions of the main system with the ancilla. In other words:
\begin{equation}
t_I \ll t_E \approx dt.
\end{equation}

In first order perturbation theory, equation (\ref{psif}) becomes
\begin{eqnarray}
\ket{\Psi_f} &=& \sum_{m,n}\bra{\tilde{\phi}_m}V_E\ket{\phi_n}\braket{\phi_n|\psi_0}\ket{\tilde{\phi}_m}\otimes\ket{\lambda\delta t_I E_{m,n}}\nonumber \\
&=& \sum_{n} \braket{\phi_n|\psi_0}\ket{\tilde{\phi}_n} \otimes \ket{\lambda\delta t_I \dot{E}_n dt} - \nonumber
\end{eqnarray}
\begin{equation}
-  dt \sum_{m,n}\dfrac{\bra{\phi_m}\dot{H}\ket{\phi_n}}{E_{m}-E_n}\braket{\phi_n|\psi_0}\ket{\tilde{\phi}_m} \otimes \ket{\lambda\delta t_I (E_{m,n} + \dot{E}_m dt)}.\label{psifinf}
\end{equation}
By analyzing equation (\ref{psifinf}), one can see that the probability of measuring the ancilla in a position that implies a quantum jump (one of the positions given by $\ket{\lambda\delta t_I (E_{m,n} + \dot{E}_m dt)}$) is proportional to $dt^2$ and can, therefore, be disregarded.

\subsection{Open System}

If the system of interest is in contact with a thermal bath, the measures of the variation of energy still must be made in a time interval $dt$ such that $H(t)$ may be considered linear. However, if one wishes not to disrupt the dynamics of the system, this interval must be much larger than the correlation time $\tau$ of the bath. If such an interval exists, during that interval, the system will evolve following a Lindblad master equation
\begin{equation}
\dot{\rho} = -\dfrac{i}{\hbar}[\rho ,H] + \sum_n \gamma_n\mathcal{L}(\sigma_n)\rho,
\end{equation}
where the Lindblad operators $\mathcal{L}(\sigma_n)\rho$ are given by
\begin{equation}
\mathcal{L}(\sigma_n)\rho = \sigma_n \rho \sigma^\dagger_n -\dfrac{1}{2}\{ \rho,\sigma^\dagger_n\sigma_n \}.
\end{equation}

In order to understand the effect of this non-unitary evolution on the results of the proposed measurement, let us consider the simple case of a two-level system with constant Hamiltonian $H(t) = H(t_0) = E_+ \ket{e} \bra{e} + E_-\ket{g}\bra{g}$ in contact with a thermal bath. The system evolves following the master equation
\begin{equation}
\dot{\rho} = -\dfrac{i}{\hbar}[\rho ,H] + \gamma_+\mathcal{L}(\sigma_+)\rho + \gamma_-\mathcal{L}(\sigma_-)\rho,
\end{equation}
where $\sigma_+ = \ket{e}\bra{g}$ and $\sigma_- = \sigma_+^\dagger$. If initialised in its ground state $\rho_0 = \ket{g}\bra{g}$, the state of the system after the evolution time $dt$, is given by
\begin{equation}
\rho_f = (1-\gamma_+ dt)\ket{g}\bra{g} + \gamma_+ dt\ket{e}\bra{e}
\end{equation}
If a measurement on the energy basis \{$\ket{e},\ket{g}$\} is made after the evolution, the probability to measure the system in state $\ket{e}$ is now proportional to $dt$ and must not be disregarded. A quantum jump occurs whenever the system absorbs a photon from, or emits a photon to the reservoir, meaning that there is an exchange of heat between the system and the bath. Similarly, when no jump occurs, the energy variation of the system is due solely to the action of an external force, resulting, therefore, from work performed on the system.
This definition becomes more precise as the probability of more than one jump in each energy monitoring step becomes negligible. This imposes limitations on the monitoring time interval that should be much smaller than the decay time, $dt \ll \frac{1}{\gamma}$.

\section{Imperfect Measures}

If the states prepared in the ancilla through the interaction with the system are completely distinguishable, i.e. orthogonal among themselves, then the measurements explained above provide the exact internal energy variation of the system in each time interval $dt$. However, in many situations, the measurement apparatus that includes the ancilla will be a continuous variable system and full distinguishability will only be achievable as a limit. 

Take, for example, a harmonic oscillator as the ancilla, initially prepared in its vacuum state. In this case, the measurement protocol drives the apparatus to energy dependent coherent states $|\alpha = \lambda t_I E_{m,n}\rangle$, which are never orthogonal among each other. That means that, in principle, there will always be some error associated to each measurement. The goal is to obtain the value of $E_{m,n}$ by performing a measurement on $|\alpha\rangle$. Assuming $\lambda$ real with no loss of generality, note that by measuring the (adimensional) position $\hat{X}$ of the harmonic oscillator, one has the probability $P(x)$
\begin{equation}
P(x)= \left| \braket{x|\alpha}\right|^2 = \dfrac{1}{\sqrt{2\pi \sigma^2}}e^{-\dfrac{1}{\sigma^2}\left(x-\alpha\right)^2}. \label{px}
\end{equation}
of finding the ancilla at the position $x$. For convenience, the standard deviation $\sigma$ will be considered to be equal to 1. It is important to remember that, from equation (\ref{psif}), to measure the ancilla at the position $x_0$ means to measure a variation of internal energy $\Delta E$ equals to $\Lambda x_0$ where $\Lambda = \lambda \delta t_I$ depends on the $\mathcal{S}-{\mathcal{A}}$ interaction time and the strength of their coupling. The probability distribution $P(E)$ of measuring $E$ for a coherent state $\ket{\Lambda E_0}$ is given by
 \begin{equation} 
 P(E)= \dfrac{\Lambda}{\sqrt{2\pi}}e^{-\Lambda^2\left(E-E_0\right)^2}.\label{fqual}
 \end{equation}
It can be seen from equation (\ref{fqual}) that the standard deviation for the energy measurement is equal to $\Lambda^{-1}$. Since the value of $\Lambda$ determines the precision of the measurements, we will call it the \textbf{Quality Factor}.

\section{Quality Factor}

\begin{figure*}[t]

\begin{center}
\subfloat[][$\Lambda = \dfrac{0.1}{\Delta E_0}$]{

\includegraphics[width=0.32\textwidth]{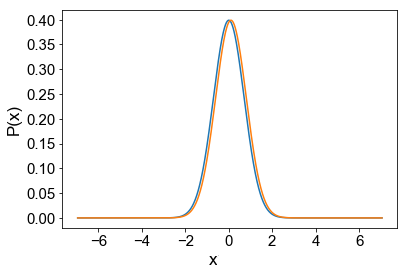}
}
\subfloat[][$\Lambda = \dfrac{1}{\Delta E_0}$]{

\includegraphics[width=0.32\textwidth]{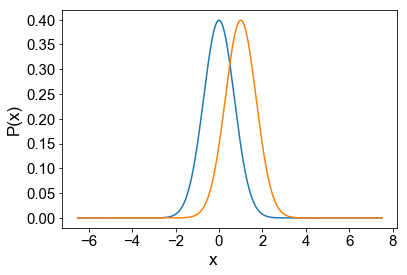}
}
\subfloat[][$\Lambda = \dfrac{10}{\Delta E_0}$]{

\includegraphics[width=0.32\textwidth]{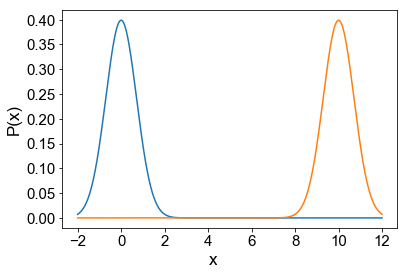}
}

\end{center}

\caption{Probability to find the ancilla at the position $x$ for various quality factors $\Lambda$. The blue curve corresponds to the case in which the energy does not change during the measure the orange curve to an energy variation of $\Delta E_0$.\label{pgauss}}

\end{figure*}

In order to better understand the effect of the quality factor $\Lambda$ on the measures here proposed, let us rewrite equation (\ref{px}) with the coherent state $\ket{\alpha} = \ket{\Lambda\Delta E_0}$, and $\sigma = 1$:
\begin{equation}
P(x)= \dfrac{1}{\sqrt{2\pi}}e^{-\left(x-\Lambda \Delta E_0 \right)^2}. \label{px2}
\end{equation}
In this form, it becomes clear that the Quality Factor $\Lambda$ defines the capacity to capture the differential energy variation $\Delta E_0$ of the system. In Figure~\ref{pgauss}, we show the displacement of the state of the Ancilla for growing values of $\Lambda$. $\Delta E_0$ is the distance between the peaks of the two curves, the unchanged one (pictured in blue) and the one shifted by the $\mathcal{S}-{\mathcal{A}}$ interaction (in orange). For small $\Lambda$'s, the overlap between both curves is large and the energy shift is practically indistinguishable. As $\Lambda$ increases, it becomes easier to observe $\Delta E_0$. For example, for $\Lambda \sim \Delta E_0^{-1}$ both peaks are clearly separated but position measurements on the Ancilla still have a reasonable probability of indicating no shift (there is still some overlap between them), whereas for a ten times larger $\Lambda$, not only are the peaks clearly distinguishable but also the probability of reading a wrong result becomes virtually zero.



Note that in order to increase $\Lambda$, one could either increase the $\mathcal{S}-{\mathcal{A}}$ interaction time $\delta_I$ or $\lambda$. As previously explained, any increase in $\delta_I$ is limited by the requirement that the system's Hamiltonian does not change significantly during the measurement procedure. On the other hand, $\lambda$ could, in principle, be increased freely without disrupting the dynamics of the system, given the form of the $\mathcal{S}-{\mathcal{A}}$ interaction, but in practice it will depend on the particular experimental setup which will, ultimately, define the quality factor of the protocol.

\section{two-level system}

\begin{figure*}[t]

\begin{center}
\includegraphics[width=0.32\textwidth]{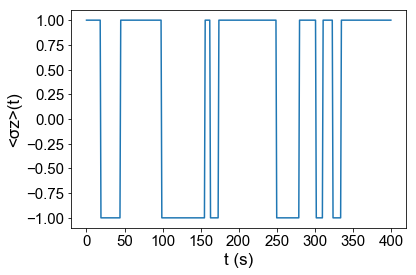}
\includegraphics[width=0.32\textwidth]{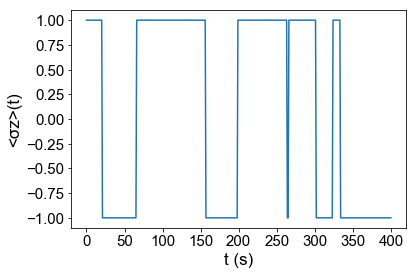}
\includegraphics[width=0.32\textwidth]{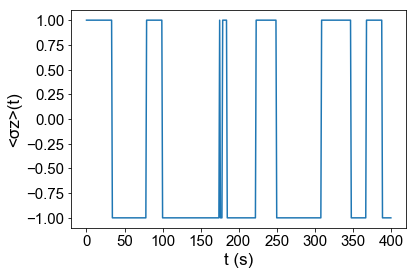}\\

\subfloat[][$\Lambda = \dfrac{1}{\hbar\omega_0}$]{

\includegraphics[width=0.32\textwidth]{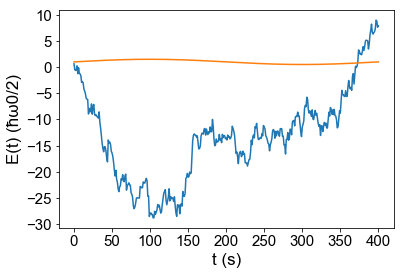}
}
\subfloat[][$\Lambda = \dfrac{20}{\hbar\omega_0}$]{

\includegraphics[width=0.32\textwidth]{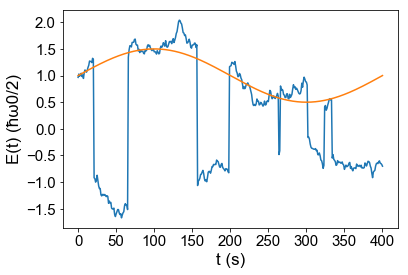}
}
\subfloat[][$\Lambda = \dfrac{100}{\hbar\omega_0}$]{

\includegraphics[width=0.32\textwidth]{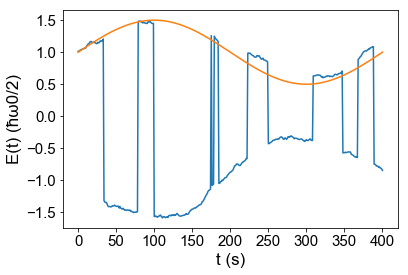}
}

\end{center}

\caption{Graphics of $\left\langle \sigma_z\right\rangle(t)$ (on the top) and $E_m(t)$ (on the bottom) for different values of quality factor. On the graphics for $E_m$(t), the blue line corresponds to the measured value for $E(t)$ and the orange line corresponds to the exact energy for the excited state $\ket{e}$.
\label{gr1}}
\end{figure*} 
 
From this point onward, we focus on the study of a particular two-level system in contact with a thermal reservoir, that is being monitored by imperfect measurements of energy as previously explained. The system of interest is one whose Hamiltonian $H(t)$ is always proportional to the Pauli matrix $\sigma_z$, but whose energy levels may vary in time $t$. In other words,  $H(t)$ is equal to
 \begin{equation}
 H(t) = \dfrac{\hbar \omega (t)}{2} \sigma_z,
 \end{equation}
where $\omega (t)$ is an unknown function of time and the Pauli matrix $\sigma_z$ is equal to
 \begin{equation}
 \sigma_z = \ket{e}\bra{e} - \ket{g}\bra{g}.
 \end{equation}
 
The system is in contact with a thermal reservoir at inverse temperature $\beta$. Thus, its evolution follows the master equation
 \begin{equation}
 \dot{\rho} = -\dfrac{i}{\hbar}[\rho ,H] + \gamma_+\mathcal{L}(\sigma_+)\rho + \gamma_-\mathcal{L}(\sigma_-)\rho,
 \end{equation}
 with $\sigma_+ = \ket{e}\bra{g}$ and $\sigma_- = \ket{g}\bra{e}$ and with $\gamma_+ = \Gamma\overline{n}(t)$ and $\gamma_- =\Gamma(\overline{n}(t)+1)$, where $\Gamma$ is the coupling strength between the system and the reservoir and $\overline{n}(t)$ is the average number of photons of the reservoir with frequency $\omega(t)$.

We will only analyse functions $\omega (t)$ that are well behaved and smooth enough to vary linearly at each measurement interval $dt$ and for which the levels $|g\rangle$ and $|e\rangle$ never cross and are always clearly split. This requires $\omega(t)$ to be always greater than an undermost value $\epsilon > 0$ ($0<\epsilon \leq \omega(t)$), and the absolute variation of $\omega (t)$ in each measurement, $|d\omega(t)|$, to be always much smaller than $\epsilon$. Beyond that, it is assumed that the state of the system at the beginning of the monitoring is known and equal either to $\ket{e}$ or $\ket{g}$. These conditions will make it possible to unravel a trajectory followed by the system, as it will be explained bellow.

 
\subsection{Detecting Effective Quantum Jumps}

In each measuring step, there are two possible outcomes: either the system stays at the same state or the reservoir induces a jump that changes the state of the system to the opposite one. Let's say the system is initially prepared in state $|e\rangle$ ($|g\rangle$) and detected at the same level after the first few measuring steps. Due to the assumption that $\omega(t)$ varies linearly in $dt$, the exact value of the system's internal energy change in each step $t_n$ will be worth $\Delta U_{e\rightarrow e} (t_n)= \dfrac{\hbar d\omega (t_n)}{2}$ ($\Delta U_{g\rightarrow g}(t_n)=-\dfrac{\hbar d\omega (t_n)}{2}$). On the other hand, if the reservoir induces a jump on the system at time $t_n$, then its exact energy variation will be worth $\Delta U_{e\rightarrow g} (t_n)= -\hbar \omega (t_n) - \dfrac{\hbar d\omega (t_n)}{2}$ ($\Delta U_{g\rightarrow e} (t_n)= \hbar \omega (t_n) + \dfrac{\hbar d\omega (t_n)}{2}$). 

Since $\omega (t) \gg |d\omega (t)|$ at all times, if the measurement's  quality factor $\Lambda$ is large enough so that the average error $\dfrac{1}{\Lambda}$ is much lower than $\hbar\omega (t)$, an effective quantum jump can easily be discerned by the measurements. In that case, a perfectly executed protocol clearly allows for the reconstruction of $\omega(t)$ from the collection of its small variations $\{d\omega(t_n)\}$.

Note that the protocol to monitor the system's energy can only distinguish between even and odd number of jumps in each interval, i.e. if the measurement interval $dt$ is not small enough to neglect the probability of two or more jumps, the protocol will bunch even number of jumps as `no-jump' and odd numbers as a single \textit{effective} jump. However, as we show bellow, this will not affect our goal of determining $H(t)$.

The exact value for each measurement is obtained when the state of the ancilla is found to be at the center of the probability distribution. The deviation between the measured value and the center of the distribution is the error of that measurement. The exact value of the change in internal energy $dE$ in each measurement relates to the exact value of the change in $\omega (t)$ in a different way for each of the cases stated previously. And so, by calling $dE_m$ the measured value for the variation of internal energy and $\omega_m$ the measured value for $\omega(t)$ at the beginning of the measurement, the measured variation of frequency $d\omega_m$ will be given by
\begin{equation}
d\omega_m = \dfrac{\left( -1\right)^{f+1}2 dE_m}{\hbar} -2\omega_m(1-\delta_{i,f}). \label{equnif}
\end{equation}
Here the initial state of the system is associated to the variable $i$ such that, if the system starts the measurement at the ground state $\ket{g}$, we have $i=0$; if it starts at the excited state $\ket{e}$, $i=1$. The final state is associated to the variable $f$ in the exact same manner.





\subsection{Determining Heat and Work}

\begin{figure*}[t]

\begin{center}
\includegraphics[width=0.32\textwidth]{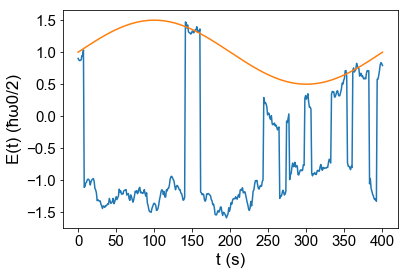}
\includegraphics[width=0.32\textwidth]{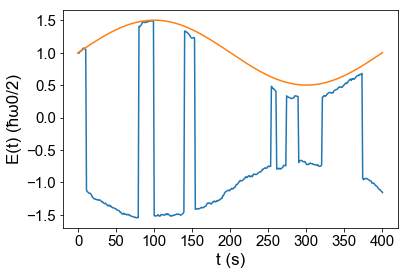}
\includegraphics[width=0.32\textwidth]{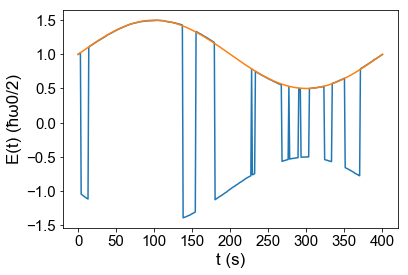}\\

\subfloat[][$\Lambda = \dfrac{20}{\hbar\omega_0}$]{

\includegraphics[width=0.32\textwidth]{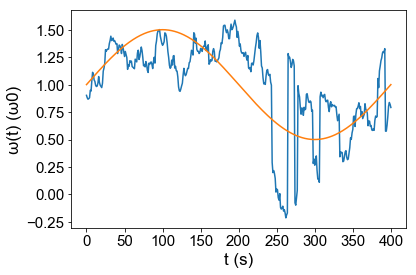}
}
\subfloat[][$\Lambda = \dfrac{100}{\hbar\omega_0}$]{

\includegraphics[width=0.32\textwidth]{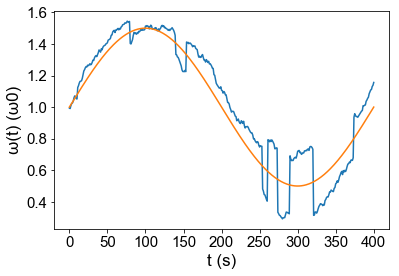}
}
\subfloat[][$\Lambda = \dfrac{1000}{\hbar\omega_0}$]{

\includegraphics[width=0.32\textwidth]{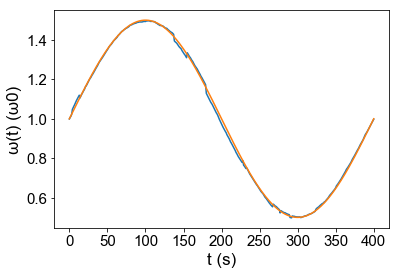}
}

\end{center}

\caption{Graphics of $E_m(t)$ (on top) and correspondent graphics of $\omega_m(t)$ (on the bottom) for various quality factors $\Lambda$. On the energy graphics, the blue curve corresponds to the measured value and the orange curve to the exact energy of level $\ket{e}$. On the frequency graphics, the blue line is the measured value and the orange line the exact value. \label{gr2}}

\end{figure*}

As in the classical case, the possible means of varying the internal energy of the system are denoted Heat and Work. Also as in the classical case, the energy exchanged between the system and the thermal reservoir will be called Heat. That energy exchange comes in the form of absorption or emission of photons by the system, and is detected by a change in its state during the measurement.

Analogously, the variation of internal energy forced upon the system by the action of an external force will be called Work. This is represented by the continuous change in the system's Hamiltonian. Therefore, whenever the system's state is unchanged during the measurement, it is said that its energy variation is entirely due to \textit{Work} performed on it.

Those definitions become exactly the same as the ones usually employed ($dQ = \text{Tr}\{ \dot{\rho}H\}dt$ and $dW = \text{Tr}\{{\rho} \dot{H}\}dt$, where $\rho$ is the system's density matrix) if our measurements are performed on time intervals sufficiently small so that the probability of the system jumping twice is negligible.

\section{Results}

The results shown in this section were obtained from a program that simulates measurements as explained previously on a two-level system that follows a randomly generated quantum trajectory. 


\begin{figure}[h]

\includegraphics[width=0.5\textwidth]{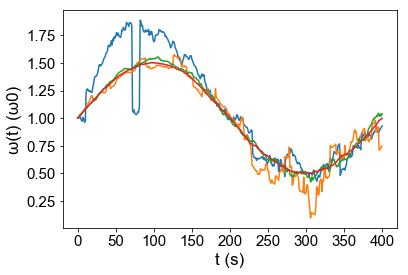}

\caption{Graphic of $\omega(t)$ with quality factor $\Lambda = \dfrac{50}{\hbar\omega_0}$. The blue line corresponds to a single realization; the orange line corresponds to the mean value between 10 realizations; the green line, the mean of 100 realizations and the red line, 1000 realizations. \label{gr5}}

\end{figure}

The program simulates a system with Hamiltonian $H(t)$ given by
\begin{equation}
H(t) = \dfrac{\hbar \omega_0}{2}\left( 1+\dfrac{1}{2}\sin( \Omega t) \right) \sigma_z,
\end{equation}
throughout one cycle.

\begin{figure*}[t]

\begin{center}
\includegraphics[width=0.32\textwidth]{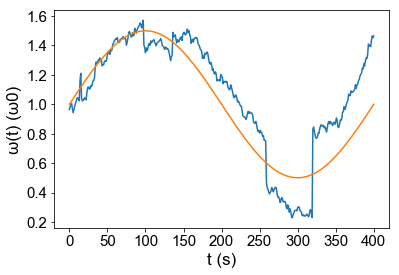}
\includegraphics[width=0.32\textwidth]{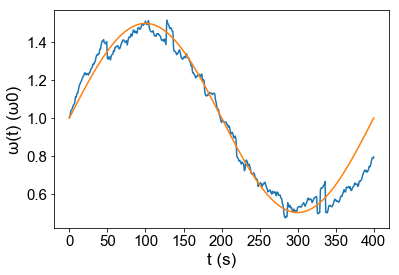}
\includegraphics[width=0.32\textwidth]{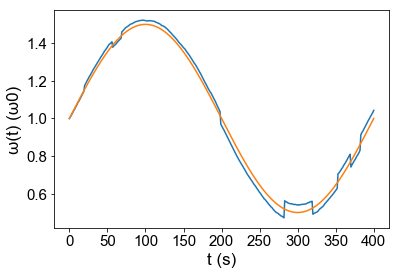}\\

\subfloat[][$\Lambda = \dfrac{50}{\hbar\omega_0}$]{

\includegraphics[width=0.32\textwidth]{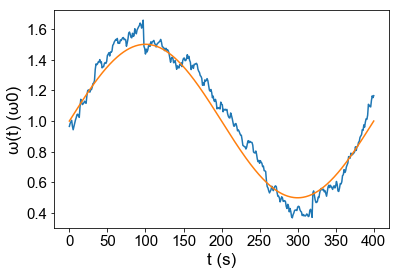}
}
\subfloat[][$\Lambda = \dfrac{100}{\hbar\omega_0}$]{

\includegraphics[width=0.32\textwidth]{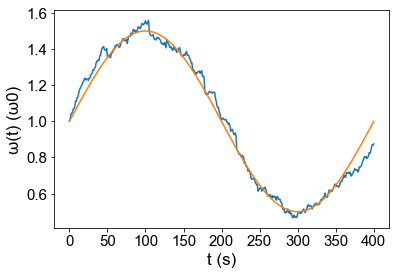}
}
\subfloat[][$\Lambda = \dfrac{1000}{\hbar\omega_0}$]{

\includegraphics[width=0.32\textwidth]{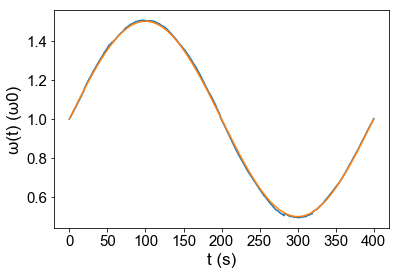}
}

\end{center}

\caption{Graphs of $\omega(t)$ made using eqs. (\ref{equnif}) on the top and (\ref{equnif2}) on the bottom. Each pair corresponds to a single simulation. In all the graphics, the blue line corresponds to the measured value and the orange line to the exact value.\label{gr3}}

\end{figure*}

\begin{figure*}[t]

\begin{center}

\subfloat[][$\Lambda = \dfrac{10^2}{\hbar\omega_0}$]{
\includegraphics[width=0.33\textwidth]{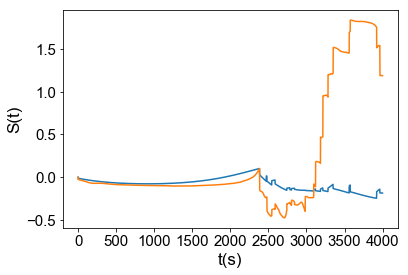}
}
\subfloat[][$\Lambda = \dfrac{10^3}{\hbar\omega_0}$]{
\includegraphics[width=0.33\textwidth]{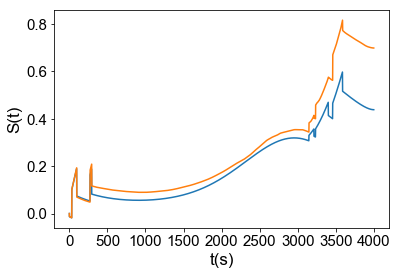}
}
\subfloat[][$\Lambda = \dfrac{10^4}{\hbar\omega_0}$]{
\includegraphics[width=0.33\textwidth]{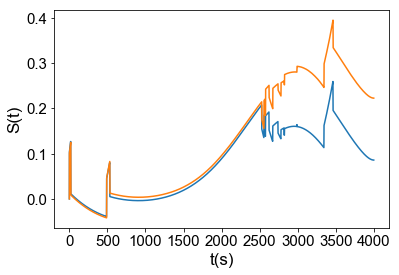}
}\\
\subfloat[][$\Lambda = \dfrac{10^5}{\hbar\omega_0}$]{
\includegraphics[width=0.33\textwidth]{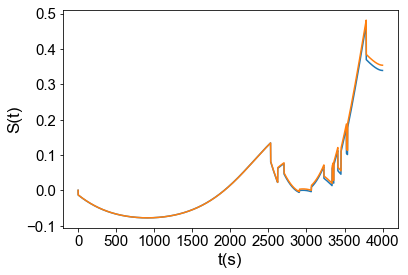}
}
\subfloat[][$\Lambda = \dfrac{10^6}{\hbar\omega_0}$]{
\includegraphics[width=0.33\textwidth]{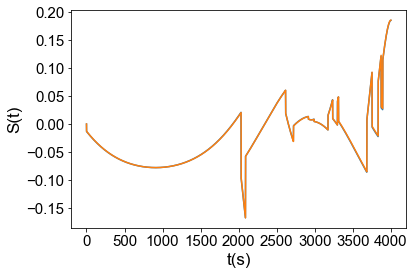}
}

\end{center}

\caption{Measured (yellow) and exact (blue) values of entropy produced as a function of time for simulations with various quality factors $\Lambda$.  \label{gr4}}

\end{figure*}

Fig.~\ref{gr1} shows $\left\langle\sigma_z\right\rangle (t)$ and the measured value for $E(t)$ for various values of $\Lambda$. We can see here how the quality factor influences the measured energy. For low quality factor  $\Lambda$, depicted in Fig.~\ref{gr1} (a), the measured energy displays a significant discrepancy with the exact one. As $\Lambda$ increases, the jumps that occur in the system are properly detected by the protocol as seen in Fig.~\ref{gr1} (b) and (c), as well as the variation of $H(t)$. In Fig.~\ref{gr2} we show the reconstruction of $\omega (t)$ as a function of the quality factor. It becomes clear that for sufficiently large values of $\Lambda$, the protocol captures with great precision the time dependence of the Hamiltonian. 

%

It is important to notice that, although measured values for $\omega (t)$ may differ vastly from the exact one for low quality factors, as can be seen in Fig.~\ref{gr2} (a) and (b), the average of $\omega (t)$ over many trajectories do tend to the exact one. This result is shown in Fig.~\ref{gr5}, where we compare a single trajectory (in blue) with the mean value of $\omega (t)$ for 10, 100 and 1000 trajectories, all made with a same low quality factor.
 
Another interesting property that can be obtained from the graphics shown in Fig.~\ref{gr2} is that, due to the particular symmetry of our Hamiltonian, there is an extra correction that can be done to improve the results. In order to understand this, one needs to note that whenever the system jumps during a measurement, the value of $\omega_m(t)$ also jumps but somehow partially compensating for the errors accumulated from previous measurements. 
In each time step, $m$, the computed value of $\omega_m(t)$ relates to the value of $E_m(t)$ as $\omega_m (t) = \dfrac{2E_m(t)}{\hbar}$ if the system is in state $\ket{e}$, and $\omega_m (t) = -\dfrac{2E_m(t)}{\hbar}$ if the system is in state $\ket{g}$. 

If the measured value for the energy immediately before a jump from $\ket{e}$ to $\ket{g}$ is equal to $E_m(t_{j_-}) = E(t_{j_-}) + \Delta E$, where $E(t_{j_-})$ is exact and $\Delta E$ is the total error accumulated by all the previous measurements, the computed value for $\omega (t_{j_-})$ will be $\omega_m(t_{j_-}) = \omega(t_{j_-}) + \Delta \omega$. Right after the jump, $E_m(t_{j_+}) = E(t_{j_-})+ dE + \Delta E + \delta E$, where $dE$ is the exact energy difference between $t_{j_-}$ and $t_{j_+}$ and $\delta E$ is the small error from a single measurement, while the computed value for $\omega$ will be given by $\omega_m(t_{j_+}) = \omega(t_{j_-}) +d\omega - \Delta \omega + \delta \omega$. This means that, by considering the actual value of $\omega (t_{j_+})$ to be $\dfrac{\omega_m (t_{j_-})+\omega_m (t_{j_+})}{2}$, the accumulated error is effectively changed to $\dfrac{d\omega}{2}$. 
This means that, instead of using Eq.~(\ref{equnif}) one should use 
\begin{equation} 
d\omega_m = \dfrac{\left( -1\right)^{f+1}2 dE_m}{\hbar}\delta_{i,j} + \left(\dfrac{ (-1)^f dE_m}{\hbar} - \omega_m(t)\right)\left(1-\delta_{i,j}\right)\label{equnif2} 
\end{equation}
It is important to note, however that this method only works because our system's Hamiltonian is always proportional to $\sigma_z$. In other systems, it may not be possible to obtain such a method to reduce the error. 

The next result shown in Fig.~\ref{gr3} is the comparison between  Eqs.~(\ref{equnif}), and (\ref{equnif2}) in order to obtain $\omega_m(t)$. It can be clearly seen that, in all the cases shown here, this method lowers the error, by making the measured (blue) curve much closer to the exact (orange) one.

\section{Entropy Production}

So far, the measurements here proposed have been used to obtain quantities related to the First Law of Thermodynamics, such as the work performed on the system, the heat it exchanges with an external reservoir, the variation of its internal energy as well as to reconstruct the time dependence of its Hamiltonian. We proceed now to calculate the entropy produced in the trajectories and to connect our results with the Second Law of Thermodynamics.

By definition, the entropy produced during a trajectory $\overrightarrow{\Sigma}$ is given by
\begin{equation}
\Delta S = \dfrac{P(\overrightarrow{\Sigma})}{\tilde{P} (\overleftarrow{\Sigma})}
\end{equation}
where $ P(\overrightarrow{\Sigma})$ is the probability that the trajectory $\overrightarrow{\Sigma}$ takes place, and $\tilde{P} (\overleftarrow{\Sigma})$ is the probability for the reverse trajectory $\overleftarrow{\Sigma}$ to take  place.

Determining both those probabilities is a simple task whenever one knows $H(t)$ and the exact trajectory followed by the system, i.e. the specific jump and no-jump sequence of $\overrightarrow{\Sigma}$.


When comparing the results for the determination of the time variation of $H(t)$ in Figure \ref{gr3} versus the reconstruction of the entropy production, depicted in Figure \ref{gr4}, one can see that the entropy production is much more sensible to the quality factor. This means that, in order to obtain a good measurement for the production of entropy, the quality factor needs to be about two orders of magnitude higher than to obtain a precise measure of the energy, and, consequently, the time dependence of $\omega(t)$. This is a direct result of the fact that loss of information due to imperfect measurements is actually another form of entropy production, classical for sure, but still adding to the one inherent to the thermodynamical process. Improving the quality factor minimizes this effect.

\begin{figure}[h]
\includegraphics[width=0.5\textwidth]{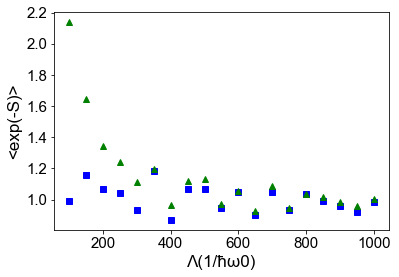}

\caption{Mean value for $\left\langle\exp\left(-\Delta S\right)\right\rangle$. The green triangles represent the mean for 1000 measured values, the blue squares represent the exact mean value for the same 1000 trajectories. For each value of $\Lambda$ there were randomized 1000 different trajectories. \label{gr6}}
\end{figure}

This can be seen if one plots the verification of the central Fluctuation Theorem which states that $\left\langle\exp\left(-\Delta S\right)\right\rangle = 1$. In Figure \ref{gr6}, one can see that for low quality factors, the Theorem applied to individual trajectories, depicted in green, seems to be violated, only converging to the proper value when $\Lambda \rightarrow \infty$. On the other hand, when the average over trajectories is taken before the entropy production is calculated, depicted in blue, then the convergence to the proper value is much faster. This behaviour results from the random and symmetric nature of the errors induced by imperfect measurements. They can affect the readout of each trajectory significantly therefore inducing a very wrong calculation of the entropy produced in it, whereas the convex sum of the trajectories tend to average out these errors, at least for the values of $\Lambda$ for which the states of the Ancilla are clearly distinguishable ($\Lambda \gg \frac{1}{\hbar \omega}$) and for a large enough number of trajectories.

Another way to approach this is to add a correction term to the measured entropy production that takes into account the information loss, much in the same way as it was carried out in~\cite{elouard2017probing}. In this case, the fluctuations of the measured entropy $\Delta S_m$ will obey the following relation $\left\langle\exp\left(-\left(\Delta S_m + \sigma_\Lambda\right)\right)\right\rangle = 1$, where the trajectory-dependent correction term $\sigma_\Lambda$ tends to zero as $\Lambda$ tends to infinity. As shown in~\cite{elouard2017probing}, since $\sigma_\Lambda$ refers to lost classical information, it can be obtained, in general, by generating correction intermediate trajectories as long as one knows the basic nature of the randomness brought by the imperfect measurements. Also note that, the apparent violation of the Central Fluctuation Theorem provides an useful strategy to infer the quality factor $\Lambda$ if this quantity is unknown.


\section{Conclusion}

We presented a method to continually monitor the energy of a quantum system and we connected our results to the thermodynamics of closed and open quantum system dynamics. We also presented a particular case in which we show that the time-dependence of the Hamiltonian and the entropy produced during the dynamics can be obtained from analyzing the energy of the system, even with imperfect measurements. Some of the results obtained were clearly dependent on the assumption that the Hamiltonian of the system of interest was always proportional to the Pauli matrix $\sigma_z$, however, the scheme developed here can also be used for other types of evolutions. Also note that the tool used in our scheme is experimentally feasible and was already implemented in an ensemble of cold atoms controlled by an atom chip \cite{roncaglia2014work}.

\bibliographystyle{ieeetr}
\bibliography{ref}

\end{document}